\documentstyle[epsf]{mn2e}

\title[The missing link: a LMXB in M31 seen as a ULX]
{The missing link: a low mass X-ray binary in M31 seen as an ultraluminous X-ray source}

\author[M.Middleton et al.]
{Matthew J. Middleton, Andrew D. Sutton, Timothy P. Roberts, Floyd E. Jackson \newauthor and Chris Done\\
Department of Physics, University of Durham, South Road, Durham
DH1 3LE,
UK\\
}

\pagerange{\pageref{firstpage}--\pageref{lastpage}} \pubyear{2011}
\long\def\symbolfootnote[#1]#2{\begingroup\def\thefootnote{\fnsymbol{footnote}}\footnote[#1]{#2}\endgroup} 
\def\ga{\mathrel{\hbox{\rlap{\hbox{\lower4pt\hbox{$\sim$}}}{\raise2pt\hbox{$>$}}
}}}

\begin{document}

\topmargin = -0.5cm

\maketitle

\label{firstpage}

\begin{abstract}

A new, transient ultraluminous X-ray source (ULX) was recently
discovered by {\it Chandra} in M31 with a luminosity at $\sim$
5$\times$10$^{39}$ erg s$^{-1}$.  Here we analyse a series of five
subsequent {\it XMM-Newton} observations. These show a steady decline
in X-ray luminosity over 1.5 months, from $1.8-0.6\times 10^{39}$ erg
s$^{-1}$, giving an observed {\it e}-fold timescale of $\sim
40$~days. This is similar to the decay timescales seen in multiple
soft X-ray transients in our own Galaxy, supporting the interpretation
of this ULX as a stellar mass black hole in a low-mass X-ray binary
(LMXB), accreting at super Eddington rates. This is further supported
by the lack of detection of an O/B star in quiescence and the spectral
behaviour of the {\it XMM-Newton} data being dominated by a disc-like
component rather than the power-law expected from a sub-Eddington
intermediate-mass black hole.
 
These data give the best sequence of high Eddington fraction spectra
ever assembled due to the combination of low absorption column to M31
and well calibrated bandpass down to 0.3~keV of {\it XMM-Newton} in
full frame mode. The spectra can be roughly described by our best
current disc model, {\sc bhspec}, assuming a 10~M$_\odot$ black hole
with best fit spin $\sim 0.4$, declining from L/L$_{\rm Edd}$ =
0.75--0.25. However, the data are better described by a two component
model, where the disc emission is significantly affected by advection,
and with an additional low temperature Comptonisation component at
high energies which becomes more important at high luminosities. This
could simply indicate the limitations of our current disc models,
though changes in the energy-dependent variability also weakly
supports a two component interpretation of the data.

Irrespective of the detailed interpretation of the spectral
properties, these data support the presence of accretion onto a
stellar mass black hole in a LMXB accreting in the
Eddington-regime. This allows an unambiguous connection of this
object, and, by extension, similar low luminosity ULXs, to `standard'
X-ray binaries.

\end{abstract}

\section{Introduction}

\begin{figure*}
\begin{center}
\begin{tabular}{l}
 \epsfxsize=13cm \epsfbox{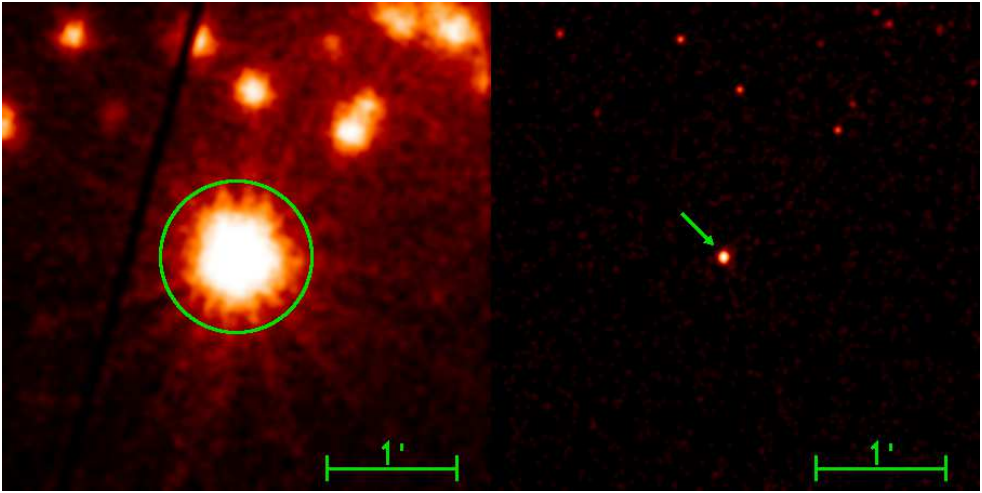}
\end{tabular}
\end{center}
\caption{{\it Left: XMM-Newton} EPIC PN image of M31 ULX-1 with the
35'' extraction region overlaid. {\it Right: Chandra} HRC-I image of
the source taken eleven days earlier. Through performing relative astrometry between the {\it Chandra} and {\it XMM-Newton} fields, we are confident that these are images of the same
source.}
\label{fig:l}
\end{figure*}

The broad population of X-ray sources in our own Galaxy includes,
amongst others, X-ray binaries (XRBs) believed to be powered by
accretion of material onto a compact object, usually a neutron star or black hole (BH). These dominate the X-ray
sky of the Galaxy and are important objects to study as they provide
high quality data, allowing an insight into the nature of accretion. In
the case of low-mass companion systems (LMXBs) the X-ray emission
characteristics are well described by the presence of an optically
thick, geometrically thin accretion disc (Shakura \& Sunyaev 1973)
and/or Comptonisation of the disc emission in a hot plasma (Titarchuk
1994). However, as the luminosity of these systems approaches their
Eddington limit our understanding breaks down. Proposed models have
included the production of large-scale winds driven by the large
radiation pressure (Poutanen et al. 2007; Kajava \& Poutanen 2009;
King 2004) whilst others favour the disc structure becoming
radiatively inefficient and instead dominated by optically thick
advection processes (Abramowicz et al. 1988; Mineshige et
al. 2000). Although a small number of XRBs have been inferred to
accrete at these high rates (V404 Cyg: {\.Z}ycki, Done \& Smith 1999;
GRS~1915+105: McClintock \& Remillard 2006; V4641: Revnivtsev et
al. 2002; Cir X-1: Done \& Gierli{\'n}ski 2003), the large intervening
column of material within our own Galaxy prevents a detailed analysis
of the behaviour of the disc. We cannot even look to the active centre
of distant galaxies (AGN) for answers, as mass scaling places the disc
emission firmly in the unobservable UV (although see Czerny et
al. 2011; Done et al. 2011).

\begin{table*}
\begin{center}
\begin{minipage}{152mm}
\bigskip
\caption{Observations of M31 ULX-1}
\begin{tabular}{l|c|c|c|c|c|c}
  \hline

Instrument & OBSID & obs. date & useful exposure (ks)& ${\it f}_{\rm x}$ & count rate \\
 &      &   &  & ($\times 10^{-11} \rm ~erg~cm^{-2}~s^{-1}$)
& (ct s$^{-1}$) \\
   \hline
{\it Chandra HRC-I}   &  10886      & 17-12-2009 &  19.0 & 6.40$^*$ & 3.3  \\
{\it Swift XRT}       &  29762      & 22-12-2009 &  4.4  & 2.40$^*$ & 0.7  \\
{\it XMM-Newton EPIC} &  0600660201 & 28-12-2009 &  14.5 & 2.52     & 4.2  \\
{\it XMM-Newton EPIC} &  0600660301 & 07-01-2010 &  13.3 & 1.87     & 4.6  \\
{\it XMM-Newton EPIC} &  0600660401 & 15-01-2010 &  6.4  & 1.37     & 3.6  \\
{\it XMM-Newton EPIC} &  0600660501 & 25-01-2010 &  10.0 & 1.14     & 2.9  \\
{\it XMM-Newton EPIC} &  0600660601 & 02-02-2010 &  9.7  & 0.86     & 2.4  \\
   \hline

\end{tabular}
Notes: Observations of M31 ULX-1 by various missions over a
period of $\sim$ 1.5 months. The useful exposure is the
length of the observation following removal of background flares in the {\it
Swift} XRT and {\it XMM-Newton} EPIC cameras (PN value given)
respectively. ${\it f}_{\rm x}$ is the unabsorbed, integrated flux
between 0.2--10~keV (from applying a simple continuum model to the data) and count rate is given for the PN in the {\it
XMM-Newton} observations. Fluxes denoted by an asterisk ($^*$) are taken from the literature values (Henze et al. 2009). 
\end{minipage} 

\end{center}
\end{table*}

The proximity of our nearest neighbour galaxy, M31, has allowed the
latest generation of observatories to detect (and often resolve) the
X-ray bright menagerie of objects it plays host to. The range of
objects in M31 is analogous to the population of X-ray sources we see
in our own Galaxy (e.g. Pietsch et al. 2009). However, unlike our
Galaxy, observations of the M31 population is not hindered by the
large absorbing column of the Galactic disc (see Warwick et al. 2011) and the distance to the source has a much smaller relative uncertainty, improving the constraints on the X-ray luminosity.

One remarkable object in M31 has been seen to emit at very high
luminosities based on the estimated distance of 0.7-0.8 Mpc (see
Vilardell et al. 2010; Tanaka et al. 2010 etc.). CXOM31
J004253.1+411422 (M31 ULX-1 hereafter) was first detected by the high
resolution camera (HRC-I) on-board NASA's {\it Chandra} mission at a
count rate of 3.28 $\pm$ 0.04 ct $s^{-1}$ (Henze et al. 2009, ATEL
\#2356). The follow-up {\it Swift} observation (OBSID 29762) placed a
rough limit on the flux of the object in the {\it Chandra} observation
at $\sim$5$\times$10$^{39}$erg s$^{-1}$. Due to its luminosity ($>$
10$^{39}$erg s$^{-1}$) and distance from the centre of M31 ($>$ 2'),
this is a {\it bona fide} ultra-luminous X-ray source (ULX, see Roberts
2007). A sequence of several {\it XMM-Newton} observations, spaced at
7-10 day intervals, followed this bright detection and have captured
the emission characteristics of the source over a period of $\sim$1
month (see also Kaur et al. 2011).

Assuming that the spectral behaviour is similar to that of other
low-luminosity ULXs (e.g. Middleton, Sutton \& Roberts 2011; Gladstone, Roberts \& Done 2009; Weng et al. 2009; Dubus, Charles \& Long 2004; La Parola
et al. 2003; Parmar et al. 2001; Takano et al. 1994) then we expect
there to be a significant disc blackbody component to the
emission. Whilst ULXs are generally persistent in nature (although see
e.g. Ghosh et al. 2006, Sivakoff et al. 2008), this object is clearly transient and, on this
basis, has much more in common with the larger population of Galactic
XRBs. These observations may therefore provide an unobscured
description of accretion through a disc at high mass accretion
rates. In this paper we will describe the temporal and spectral
properties obtained from the high quality {\it XMM-Newton} data as the
source dims from its peak observed brightness.

\section{Data analysis}

In total, M31 ULX-1 has been observed 7 times in outburst; initially
it was detected using the {\it Chandra} HRC-I, followed by a {\it
Swift} target of opportunity (ToO) and then 5 {\it XMM-Newton}
observations. The details of each of these are given in Table 1. To
confirm that we are viewing the same object in all the observations we
extracted the HRC-I data of the field when the source is bright
(Fig. 1) and obtained the positions of all the point sources using
{\sc celldetect} in {\sc ciao v 4.1}. We then used {\sc edetect} in
{\sc sas v10} to obtain the analogous source positions in the EPIC PN
field from the observation chronologically closest to the {\it
Chandra} observation (OBSID 0600660201: Fig. 1). The positions of
several bright sources common to both fields were then used to
constrain their relative astrometry. The position of the ULX (as determined by the high resolution HRC-I of {\it Chandra}) at  R.A.(J2000) = 00:42:53.15, Dec.(J2000) = +41:14:22.9 (with a typical error of $\sim$0.5'') was
consistent (within errors) in both fields, confirming it as a single
transient object viewed at different epochs.

For each of the {\it XMM-Newton} detections we extracted the event files
using {\sc sas v10} and filtered the products for standard patterns
($<$=12 for MOS, $<$= 4 for PN) and flags (=0 for spectral and timing
products). From these we extracted spectral and timing properties from
circular, 35'' radius source and background regions from the same chip
avoiding other sources in the field. We also filtered for hard (10--15~keV) background flaring intervals based on data from the entire chip,
leaving the useful exposures given in Table 1.

In the first observation there was partial pile-up in the PN which we
corrected for by removal of a 5'' radius region around the centroid of
the PSF, checking that the observed patterns were within the error
bounds of the model patterns using the tool {\sc epatplot}. We found
significant pile-up in the the first two MOS2 observations and, due to
requiring large centroids for removal, we instead ignore these
datasets.

We extracted spectral products using {\sc xselect}, obtaining background-subtracted spectra and responses. We fit these using {\sc comptt} to
phenomenologically describe the continuum, with a neutral absorption
column\footnotemark\footnotetext{with a lower limit based on the
Galactic line-of-sight to M31 of $\sim$6.7$\times$10$^{20}$cm$^{-2}$
(Dickey \& Lockman 1990)}. This describes the spectra well
($\chi^2$ = 2809/2789) for electron temperatures ranging from
0.94--0.78~keV, i.e. all spectra have a rollover well below 10~keV.
We use this model to derive the unabsorbed 0.2--10~keV
fluxes reported in Table 1.

We plot the unabsorbed flux as a function of time in Fig 2. There is a
clear exponential decay, similar to that often seen in X-ray novae
which can be modelled as $f(t)$ = $f_{0}\times$e$^{-t/\tau}$ where $f$
is the flux after $t$ days, $f_{0}$ is the initial flux and $\tau$ is
the {\it e}-fold time. Given that the {\it Chandra} flux is only a rough
estimate (Henze et al. 2009; ATEL \#2356) we fit the full dataset
(blue dashed line in Fig. 4) and the dataset without this point (red
dot-dashed line) with a linear model in natural log space. These give {\it e}-fold times
between 29 and 39 days respectively, consistent with that observed for
X-ray novae in our own Galaxy (Shahbaz et al. 1998).


\begin{figure*}
\begin{center}
\begin{tabular}{l}
 \epsfxsize=15cm \epsfbox{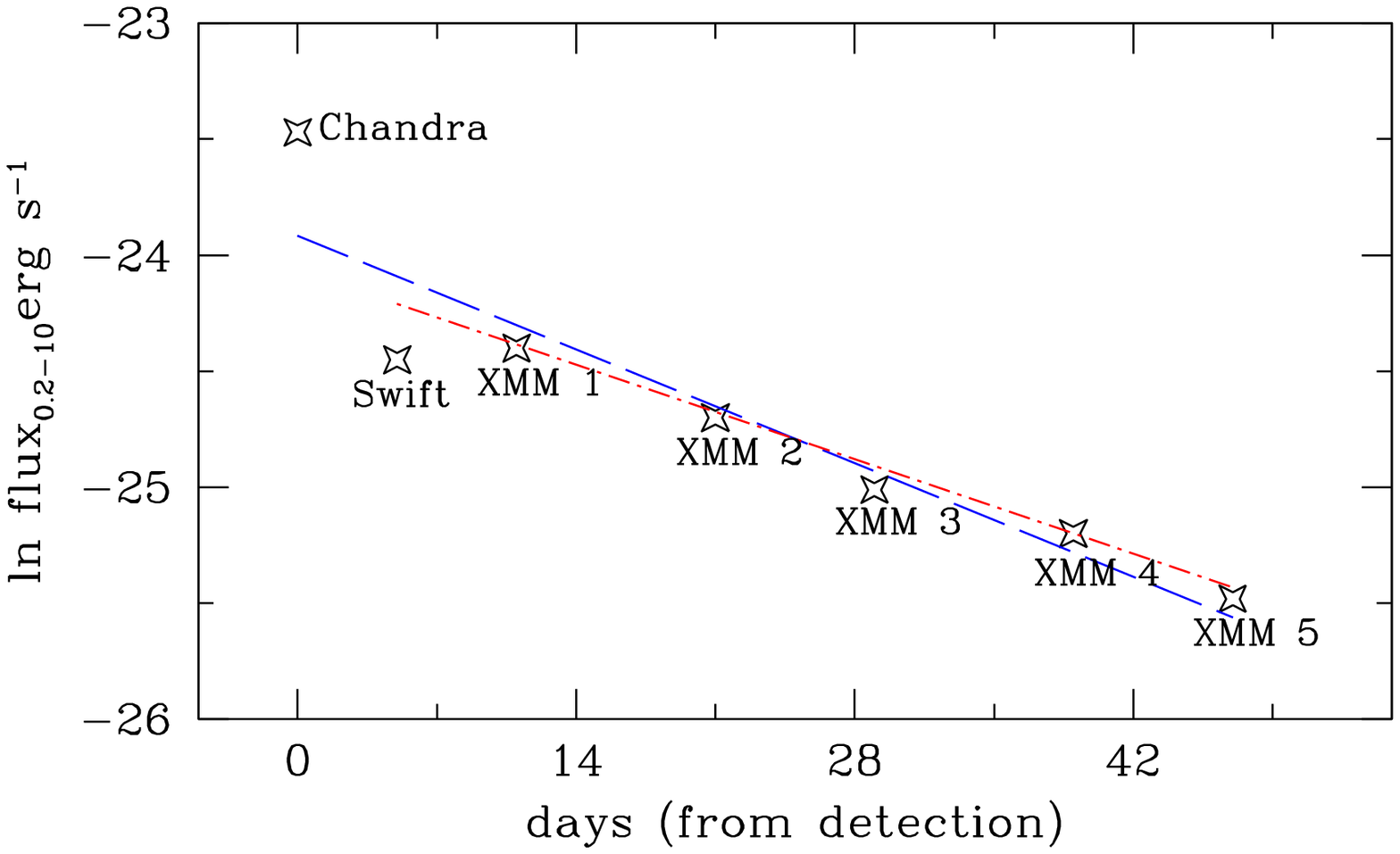}
\end{tabular}
\end{center}
\caption{Unabsorbed 0.2--10~keV fluxes of M31 ULX-1 from observations
taken with {\it Chandra}, {\it Swift} and {\it XMM-Newton} (see Table
1 for details of each of these). We fit the data with a ln-linearly
decreasing trend and find that, by including the {\it Chandra}
detection (blue dashed line), we obtain an {\it e}-fold time of 29
days and excluding it (red dot-dashed line), an {\it e}-fold time of
39 days. These derive from spline fitting in {\sc qdp} and have
variances of 0.36 and 0.03 with and without the Chandra flux point
respectively. Both of these are consistent with the observed {\it e}-fold
times of X-ray novae (e.g. Mineshige, Yamasaki \& Ishizaka 1993).}
\label{fig:l}
\end{figure*}

\section{X-ray spectra}

The {\it XMM-Newton} observations range from L$_{\rm x}$=1.8-0.6$\times10^{39}$ ergs s$^{-1}$, so only the first few datasets
meet the ULX criteria of $>$10$^{39}$ ergs s$^{-1}$. These
luminosities and heavily curved spectra (see previous section) clearly
relate to standard stellar mass BHs accreting at close to the
Eddington limit. The alternative explanation of an intermediate-mass
BH (IMBH: Colbert \& Mushotzky 1999), with 10$^{2-5}$~M$_{\odot}$,
would instead have $L/L_{Edd}\ll 0.1$ where the spectrum would be
expected to be dominated by a hard power-law extending to $\sim
100$~keV. This is plainly inconsistent with the observed rollover in
the data below $\sim$10~keV.  However, the {\it Chandra} discovery
luminosity is $\sim$ 5$\times 10^{39}$ ergs s$^{-1}$, so this source
is clearly connected to the ULX regime.

Ruling out the presence of an IMBH means that the compact object
driving the accretion is most likely a stellar mass BH. Assuming that
the BH population is similar to that of our own Galaxy then we could
be observing accretion onto a BH of mass $\sim$5--20~M$_{\odot}$ at
rates of $L/L_{Edd}\sim 0.2-2$. XRBs accreting at similar rates have
spectra which are dominated by a hot accretion disc, showing
temperatures of $\sim 1$~keV for luminosities up to Eddington
(e.g. McClintock \& Remillard 2006). In these disc dominated states,
the maximum disc temperature and total luminosity change together such
that $L\propto T^4$, indicating a constant inner emitting area, as
predicted by the existence of the innermost stable circular orbit in general relativity
(see e.g. Done, Gierli{\'n}ski \& Kubota 2007).

\begin{table*}
\begin{center}
\begin{minipage}{135mm}
\bigskip
\caption{Best fitting spectral parameters.}

\begin{tabular}{l|c|c|c|c|c}
\hline

Observation & 1 & 2  & 3 & 4 & 5 \\
\hline
\hline
\multicolumn{6}{|c|}{\sc tbabs*bhspec}\\
\hline
\hline
\multicolumn{6}{|c|}{$n_{\rm H}$ = 0.067$\times$10$^{22}$cm$^{-2}$}\\
\multicolumn{6}{|c|}{cos($i^{\circ}$) = 0.87 $_{-0.02}^{+0.04}$ }\\
\multicolumn{6}{|c|}{$a_{*}$  = 0.36 $_{-0.11}^{+0.10}$ }\\
\multicolumn{6}{|c|}{BH mass (M$_{\odot}$) = 10$^{*}$}\\
log $L/L_{\rm Edd}$  &   -0.133 $_{-0.026}^{+0.016}$ & -0.283 $_{-0.011}^{+0.015}$ &  -0.418 $_{-0.013}^{+0.016}$ &  -0.521 $_{-0.005}^{+0.005}$ & -0.612 $_{-0.005}^{+0.005}$\\
\\
\multicolumn{6}{|c|}{$\chi^2$ (d.o.f.) 3169.4  (2801)}\\
\multicolumn{6}{|c|}{Null P 1$\times$10$^{-6}$}\\
\hline
\multicolumn{6}{|c|}{$n_{\rm H}$ $<$ 0.068$\times$10$^{22}$cm$^{-2}$}\\
\multicolumn{6}{|c|}{cos($i^{\circ}$) = 0.86$^{*}$}\\
\multicolumn{6}{|c|}{$a_{*}$ $>$0.76}\\
\multicolumn{6}{|c|}{BH mass (M$_{\odot}$) = 14.9 $_{-0.8}^{+0.2}$}\\
log $L/L_{\rm Edd}$  & -0.257 $_{-0.005}^{+0.015}$& -0.403 $_{-0.005}^{+0.016}$& -0.537 $_{-0.006}^{+0.016}$& -0.640 $_{-0.005}^{+0.016}$& -0.731 $_{-0.006}^{+0.016}$\\
\\
\multicolumn{6}{|c|}{$\chi^2$ (d.o.f.) 3078.5  (2801)}\\
\multicolumn{6}{|c|}{Null P 2$\times$10$^{-4}$}\\
\hline
\multicolumn{6}{|c|}{$n_{\rm H}$ = 0.067$\times$10$^{22}$cm$^{-2}$}\\
\multicolumn{6}{|c|}{cos($i^{\circ}$) = 0.5$^{*}$}\\
\multicolumn{6}{|c|}{$a_{*}$  = 0.76 $_{-0.01}^{+0.01}$ }\\
\multicolumn{6}{|c|}{BH mass (M$_{\odot}$) $>$29.6}\\
log $L/L_{\rm Edd}$   & -0.381 $_{-0.003}^{+0.007}$ & -0.530
$_{-0.003}^{+0.006}$ & -0.664 $_{-0.005}^{+0.008}$ & -0.766
$_{-0.005}^{+0.007}$ & -0.857 $_{-0.005}^{+0.008}$\\
\\
\multicolumn{6}{|c|}{$\chi^2$ (d.o.f.)   3009.5(2801)}\\
\multicolumn{6}{|c|}{Null P  0.003 }\\
\hline 
\hline
\multicolumn{6}{|c|}{\sc tbabs*(diskpbb+comptt)} \\
\hline
\hline
\multicolumn{6}{|c|}{$n_{\rm H}$ 0.104 $_{-0.001}^{+0.001}$ $\times$10$^{22}$cm$^{-2}$}\\
\\
$kT_{\rm in}$ (keV)        & 0.44 $_{-0.08}^{+0.08}$ & 0.51 $_{-0.01}^{+0.01}$ & 0.67 $_{-0.03}^{+0.03}$ & 0.80 $_{-0.03}^{+0.03}$ & 0.92 $_{-0.03}^{+0.03}$ \\ 
$p$                        & 0.60 $_{-0.01}^{+0.03}$ & 0.60 $_{-0.02}^{+0.03}$ & 0.60 $_{-0.01}^{+0.02}$ & 0.56 $_{-0.01}^{+0.01}$ & 0.55 $_{-0.01}^{+0.01}$ \\
$norm$ & 7.59 & 3.69 & 1.14 & 0.37 & 0.18\\
$kT_{\rm comp,seed}$ (keV) & 0.55 $_{-0.13}^{+0.10}$ & 0.96 $_{-0.04}^{+0.04}$ & 0.61 $_{-0.16}^{+0.27}$ & $<$1.26 & ------ \\
$kT_{\rm comp}$ (keV)      & 0.98 $_{-0.10}^{+0.30}$ & 0.49 $_{-0.02}^{+0.03}$ & 0.81 $_{-0.04}^{+0.06}$ &  $<$11.73 & ------ \\ 
$\tau$                     & 11.33 $_{-5.22}^{+6.60}$ & 16.39 $_{-3.95}^{+4.09}$ & $>$19.00 & (25.50) & ------ \\ 
\\
\multicolumn{6}{|c|}{$\chi^2$ (d.o.f.) 2752.2 (2777)}\\

\multicolumn{6}{|c|}{Null P  0.73 }\\

\hline
\end{tabular}
Notes: {\sc bhspec} model: best-fitting values of the Eddington
fraction are given for spectral fitting, with fixed parameters denoted by $^{*}$. We firstly fit the data with the mass fixed
in the model at 10~M$_{\odot}$ (to be consistent with the BH
population of our own Galaxy) with the inclination left free to vary,
and subsequently with the inclination fixed at 30 and 60 degrees with
BH mass free to vary. {\sc diskpbb} model: best fitting parameters are
given for the 5 observations with the 5$^{\rm th}$ having no
constrained Compton component in the spectrum. Where upper or lower
limits are given, the lower (0.01~keV for the seed photon or electron
temperature) or upper (200 for the optical depth, $\tau$, 0.8 for the spin, a$_*$ and 30~M$_{\odot}$ for the BH mass) hard
boundaries for the model parameter have been reached. In the case of
the optical depth for observation 4, the value is given in brackets as
it is unconstrained. In all cases where there are two components, the
degenerate nature of the spectral fitting requires one best-fitting
component to be held in position whilst 90\% error bounds are placed
on the other. This is an unavoidable result of modelling such broad
continua.

\end{minipage}
\end{center}
\end{table*}

\subsection{Disc models}

The most accurate available model for disc emission is {\sc bhspec}
(Davis et al. 2005) which self-consistently determines the spectrum at
each radius by solving the vertical radiative transfer, and propagates
this to the observer through the full general relativistic
spacetime. These spectra are integrated over the entire disc, assuming
the stress free inner boundary condition, with innermost radius fixed
to the innermost stable circular orbit. 

We fit this together with neutral absorption$^1$ to all the
PN data across all of the observations simultaneously (we do not use the MOS datasets due to pile-up issues and the low effective area at high energies relative to the PN), fixing the
inner radius of the disc (set by the BH spin) and inclination to
be the same in all the datasets.  As this source looks similar to many
soft X-ray transients in our own Galaxy, we first assume a 10~M$_\odot$
BH. This gives a poor fit to the data
($\chi^2=3169/2801$) while returning a reasonable value for the spin and inclination (see Table
2). However, we obtain a better fit by allowing the BH mass to be
free. We tabulate results for $i=30^\circ$ and $60^\circ$,
with the latter giving the best fit ($\chi^2$ = 3009/2801) and a BH mass pegged at the upper limit of 30~M$_\odot$. We show the
data and residuals deconvolved with this best fit model
in Figure 3. 

The residuals indicate that the spectrum is somewhat broader than
the disc models (resulting in the poor fit quality). This is commonly seen in {\it XMM-Newton} spectra of disc
dominated XRBs (Kubota et al. 2010; Kolehmainen et al. 2011), but the
level of residuals here is about twice as large.  This could indicate
that the wider bandpass used here (down to 0.3~keV rather than the
0.7~keV lower limit from the other CCD observations of disc dominated
spectra) enhances the differences already seen between our best disc
models and the data. Alternatively, the residuals could indicate that
there is an additional component in the spectrum. XRBs commonly show a
power-law tail to high energies, so we allow for this using the {\sc
simpl} model for Comptonisation (Steiner et al. 2009), fixing the power-law photon index to 2.2 as is commonly seen in the high/soft state
(Kubota, Makishima \& Ebisawa 2001). We include this additional component in our best
fit model, but the improvement in $\chi^2$ is not significant.  The
scattered fraction is always $\le 5$\%, and the best fit disc
parameters are not significantly different.

\subsection{Disc plus low temperature Comptonisation}

As the disc approaches the Eddington limit, optically thick advection
should become important, where photons are swept along radially with
the flow rather than escaping vertically (slim discs: Abramowicz et al.
1988). These photons can then be released in the plunging region where
the flow becomes optically thin as it accelerates towards the event
horizon (S{\c a}dowski 2009). Another process which can become important is
a radiatively driven wind (Poutanen et al. 2007; Ohsuga \& Mineshige
2011). At very high luminosities we expect the wind to launch at
large radii in the disc, but at lower luminosities we may expect the wind
to be launched only from the innermost radii of the disc where the
radiation pressure is strongest. These processes are not mutually
exclusive, and it seems most likely that both advection and wind-driving affect the spectra at Eddington and beyond (Ohsuga 2007; Ohsuga et al.
2009).

We have previously suggested that a wind-dominated model can explain
both the spectral and timing characteristics of other ULXs at similar
luminosities (Gladstone et al. 2009; Middleton et
al. 2011b). The evolution of such systems is difficult to directly
observe as most are persistent. However, in this case we see a
clear  drop in luminosity allowing the predictions of the model to
be tested. In particular we expect that, as the mass accretion rate
drops, the outer photospheric radius of the wind moves inwards as the
radiation pressure decreases. The cool disc beyond the wind then has a
smaller inner radius and should appear hotter. We describe this in
{\sc xspec} by a model comprising a slim disc with the index of the
temperature profile, $p$, a free parameter, together with thermal
Comptonisation (with the seed temperature allowed to be free rather than being fixed to that of the disc) and neutral absorption$^{1}$ (in {\sc xspec:
tbabs*(diskpbb+comptt)}). Although there is no real advantage to
fitting the data simultaneously, we do so to allow a direct comparison
with the disc-only model ({\sc bhspec}).

\begin{figure*}
\begin{center}
\begin{tabular}{l}
 \epsfxsize=12cm \epsfbox{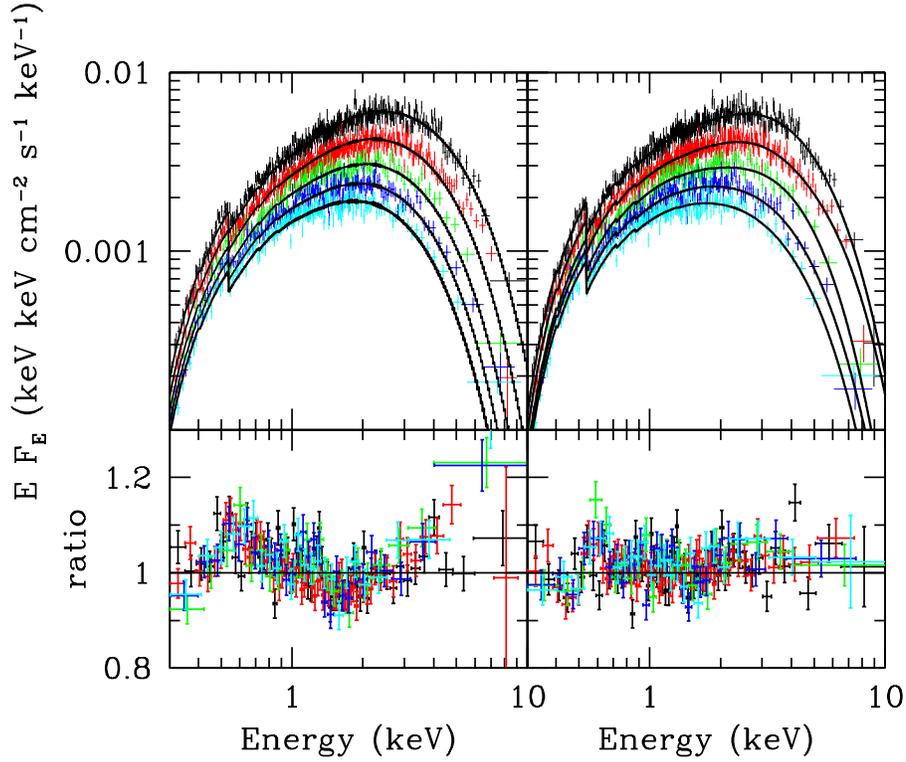}
\end{tabular}
\end{center}
\caption{Simultaneous X-ray spectral fitting for M31 ULX-1 using the
EPIC PN data across all 5 {\it XMM-Newton} observations (Obs1: black,
Obs2: red, Obs3: green, Obs4: blue and Obs5: cyan). {\it Left}: we
initially attempt to describe the data using a single disc model ({\sc
bhspec}). This can provide an acceptable fit quality but only for
neutral absorption consistent with the Galactic column to M31. This
model also suffers from systematically underestimating the flux at
soft energies as can be seen in the ratio of data to the model (lower
panel). This may indicate that the model is too simple, with the
residuals potentially indicating emission from within the optically
thin plunging region. {\it Right}: the same data but with a model of
slim disc and inner photosphere/wind production. It is clear from inspecting
the ratio to the best fitting model that the description is an
improvement. This model can also describe the data for a larger
absorption column (see Table 2).}
\label{fig:l}
\end{figure*}

The best-fitting model and associated residuals are shown together in
Figure 3 (right hand panel) with the model components highlighted in Figure 4 and the
parameters given in Table 2. We obtain an overall significant
improvement in fit quality compared to the single disc model
($\Delta\chi^2$ of 257 for 24 extra d.o.f.). The best-fitting model
parameters show that the temperature of the disc increases with
decreasing luminosity, while its radius decreases. The corresponding
Compton component peak temperature (which is a function of both seed
photon temperature and electron temperature) also increases, but the
fraction of the total luminosity carried by this drops with
luminosity. This is broadly consistent with the predicted behaviour of
a wind/photosphere launched from smaller radii as the radiation
pressure drops (Middleton et al. 2011b). A possible issue arises where
the disc at the lowest observed luminosity appears heavily advection
dominated which is inconsistent with the inferred sub-Eddington mass
accretion rate. Part of this is due to the lack of relativistic
smearing in this model which makes the disc emission artificially
narrow (Kolehmainen et al. 2011; Kubota et al. 2010).

We can obtain a rough estimate for the inner radius of the disc from
the normalisation of the {\sc diskpbb} model. This drops from an
apparent radius of 300 to 47~km assuming an inclination of $60^\circ$,
which corresponds to a true inner radius of $\sim$400 to 62~km after
correcting for colour temperature and stress free inner boundary
condition (Kubota et al. 2001). The lowest number is similar to that
derived from the {\sc bhspec} fits for a 10~M$_\odot$ BH
i.e. 71~km for $a=0.36$, whereas the larger radii seen for the disc at
higher luminosities in this model imply that the wind photosphere
dominates down to 26$R_g$. This is somewhat larger than expected given
that a 10~M$_\odot$ BH is still sub-Eddington at this point, so
should not yet power a strong wind. This would argue for a smaller
mass BH (e.g. XTE~J1650-500, Orosz et al. 2004), in order that the first few observations are super-Eddington.

\begin{figure*}
\begin{center}
\begin{tabular}{l}
 \epsfxsize=12cm \epsfbox{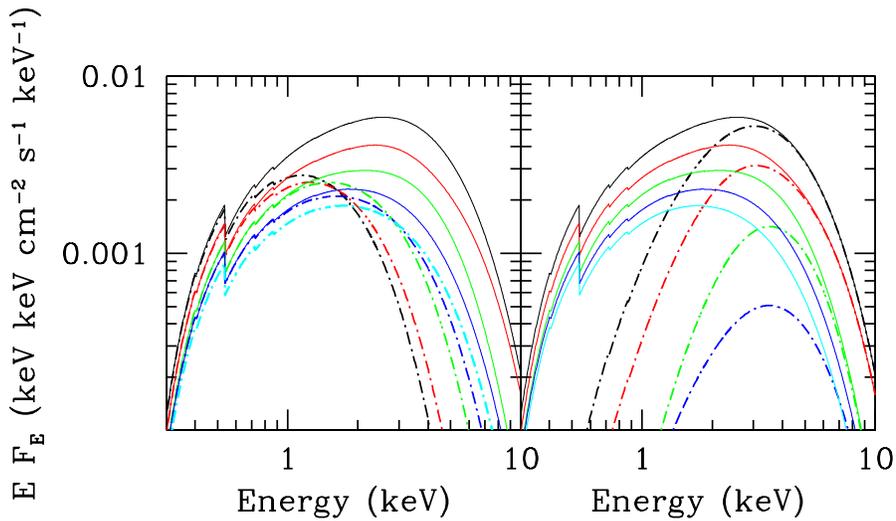}
\end{tabular}
\end{center}
\caption{{\it Left}: Contribution of the disc component to the best
fitting model of slim disk and photosphere/wind (the total model is
given as a solid line and the disc component as a dot-dashed line with
the colours corresponding to the observations in the caption of
Fig. 2). As the source dims, the model predicts an increase in
advection dominance and a broadening of the component. {\it Right}:
The contribution to the same model from the photosphere/wind (dot-dashed
line with colours as before). As the source dims, the model predicts
that this component should get hotter and less dominant as it is
driven from smaller, hotter radii.}
\label{fig:l}
\end{figure*}

\section{Short timescale variability}

The spectra alone are somewhat degenerate. As we have shown, a
description using only a single disc component (with fixed inner
radius as the luminosity declines) is statistically poor. However, the
best available models are not sufficiently well calibrated with the
expectation of residuals when compared to the data. An alternative
(and statistically better) description is a two component model, with
an advective slim disc at low energies, together with an inner-disc
wind photosphere at high energies which decreases in radius as the
luminosity declines. We now use the additional information contained
in the lightcurves to try to discriminate between these two very
different interpretations of the spectra.

We can examine the variability as a function of energy within a given
observation in several ways. The simplest of these is to obtain the
variance of a light curve in a given energy band, subtract the
expected variance of the Poisson (white) noise and normalise by the
mean count rate to get the fractional variability (excess variance, or
rms, see Edelson et al. 2002, Vaughan et al. 2003). We split the
bandpass into 4 separate energy bins, from 0.3--0.7, 0.7--1.4,
1.4--3.0 and 3.0--10.0~keV and create lightcurves in each energy
range, binned on 300~s (to provide adequate statistics for such an
analysis) from coadding the PN and MOS data (we note that, due to the
lower quality statistics, the small amount of pile-up in the first two
MOS2 datasets adds only a small constant offset to variability at hard
energies and so will have negligible effect upon the result). These
showed some weak evidence for variability. We then carefully iterated
the binning across the energy bandpass whilst ensuring adequate statistics
remained in each bin under investigation and were able to find at
least one energy range for each observation in which variability was
detected at more than 3$\sigma$. However, the upper limits on
variability in the remaining energy bins are large.

We can improve the statistics by using the energy band in which the
variability is detected as a reference lightcurve to calculate the
fractional covariance i.e. the amount of variability in the other
energy bands which is correlated with the detected variability (see
Wilkinson \& Uttley 2009). The resulting energy-dependent fractional
covariance spectra are shown in Figure 5, where the reference band
value (red point) is the fractional rms. We have attempted to obtain
as many constrained bins as possible, however, the lack of correlated
variability makes this extremely difficult leaving us with only a
small number of bins in each observation. Irrespective of this we can
quite clearly see that the reference bandpass has the highest
variability (by design), and that there is significantly less
correlated variability at other energies in observations 1, 4 and 5. Additionally, the typical energy at which this variability is seen
increases as the source declines.
We note that observation 2 does not necessarily show a lack of correlated
variability as the only other well constrained bin of excess
variance (besides the reference band) is consistent within 1 sigma with the
reference band whilst the nature of variability in observation 3 is difficult to
accurately and reliably constrain due to the relative lack of counts above 2~keV.

\begin{figure*}
\begin{center}
\begin{tabular}{l}
 \epsfxsize=13cm \epsfbox{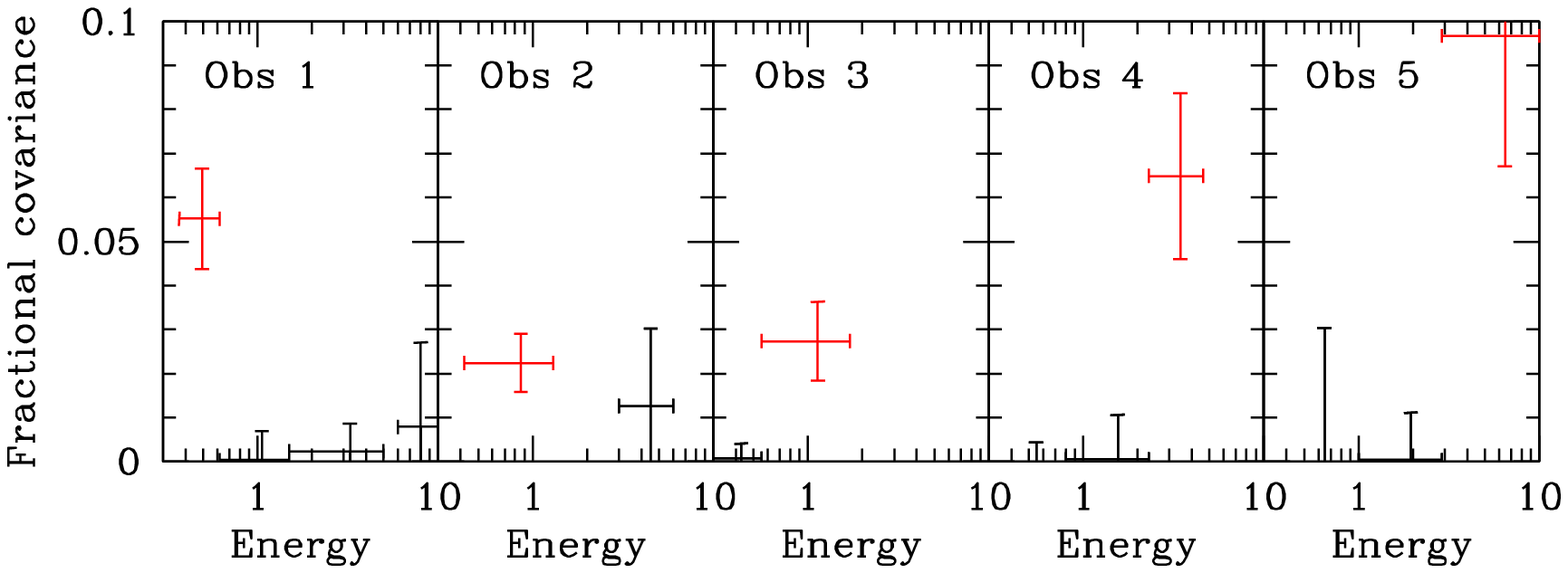}
\end{tabular}
\end{center}
\caption{Fractional covariance for each of the observations on 300~s
binning (to ensure adequate statistics) with the reference band shown
in red (for which the value is the fractional excess variance). There
is clearly a lack of evidence for correlated variability in any of the
observations, whilst the well-constrained bin of variability is seen
to increase in energy. This presents problems for our standard view of
accretion disc behaviour but can be more readily explained by our ULX
model.}
\label{fig:l}
\end{figure*}

We accept that it is possible that our iterative energy binning scheme
could distort the statistical significance of this. Indeed, testing
this approach on a similarly bright but constant source with similar
observation length (G21.5-09, OBSID: 0122700101) can produce a single
narrow bin of variability due to the inherent stochasticity of the
white noise. However, it is highly unlikely that such a process could
produce the observed energy dependence with luminosity seen in M31
ULX-1 and so we claim that there is real variability present. This
energy dependence is difficult to explain in the single disc model. A
disc dominated {\it XMM-Newton} spectrum from LMC~X-3 shows
variability on the few percent level (Kolehmainen et al. 2011, in
prep), but this is constant across the entire energy band.  Instead,
it is much easier to explain in the two component model, where the
interface between the disc and wind picks out a distinct radius and
hence temperature, and this temperature increases as the wind
photosphere radius decreases during the decline (see Fig 4). The
modulation could be due either to instabilities in the disc at this
radius, or to the wind being clumpy and stochastically covering this
inner region of the disc.


\section{Discussion \& Conclusion}

The combined spectral and timing behaviour of M31 ULX-1 allows a
unique insight into the properties of a {\it bona fide} ULX, albeit a
somewhat unusual ULX in being a transient rather than persistent
system. 

Garcia et al. (2010, ATEL \#2474) determined the presence of an optical
counterpart to the X-ray emission during the outburst, on the basis of a
$\sim 4$ ks {\it HST\/} Advanced Camera for Surveys (ACS) F435W filter
exposure taken on 2010 January 21, with an apparent magnitude of 23.8 (Figure 6, left hand panel).
This is significantly brighter than a limit we derive from a later,
post-outburst observation taken on 2010 July 20 in the same instrument, of
$m_{\rm F435W} \approx 26$ (Figure 6, right hand panel). By converting the neutral absorption column
measured from the X-ray data into an extinction using the relation of
Predehl \& Schmitt (1995), we find $A_B = 0.66$ towards the source in M31
(using the $B$ filter as a close proxy of the {\it HST\/} F435W filter).
We use this, and the magnitude limit, to derive an upper limit on the
absolute magnitude of the optical counterpart in quiescence, presumably
the emission of the secondary star in the system, of $M_{\rm F435W}
\approx +1$.  This rules out a high mass O or B star companion to the
compact object, consistent with the identification of this transient as an
LMXB.  The lack of a persistent bright optical counterpart also implies
that the detected transient optical emission is most likely from
reprocessing in the X-ray illuminated outer disc (van Paradijs 1996).

The source lightcurve also strongly supports an LMXB identification
(Fig 2). There is a clear, dramatic outburst, followed by an
exponential decay. Such outbursts are triggered by the hydrogen
ionisation instability in the outer disc, and so requires a low mass
transfer rate from the companion star. A high mass companion filling
its Roche lobe would have too large a mass transfer rate, making the
system persistent rather than transient (King 1999).

The {\it e}-fold timescale of 39 days is also very similar to that seen in
many stellar mass BH LMXB transients in our own Galaxy (Chen, Shrader
\& Livio 1997; Shabaz et al. 1998; Mineshige, Yamasaki \& Ishizaka 1993). We can use this timescale to
estimate the radius of the ionised disc (from equation 14 of Shahbaz,
Charles \& King 1998) using the peak luminosity estimate from the {\it
Chandra} observation ($\sim$5$\times$10$^{39}$ erg s$^{-1}$), the
upper limit for the {\it e}-fold time (39 days), the accretion efficiency
($\sim$0.1 for an alpha disc: Shakura \& Sunyaev 1973) and the
critical density for the disc to give an upper limit for the size
scale. This gives a radius of 2.6$\times$10$^{11}$ cm and a lower
limit for the period of the binary orbit (from Frank, King \& Raine
2002) of $>$32.8 hours. In terms of binary parameters and inferred
luminosity, this system is then qualitatively similar to 4U 1543-47
(Chen et al. 1997), reinforcing our identification of the system as a
LMXB. For a low mass star to fill its Roche lobe in such a wide binary
means that the companion is probably somewhat nuclear evolved, i.e. a
sub-giant.

\begin{figure*}
\begin{center}
\begin{tabular}{l}
 \epsfxsize=5cm \epsfbox{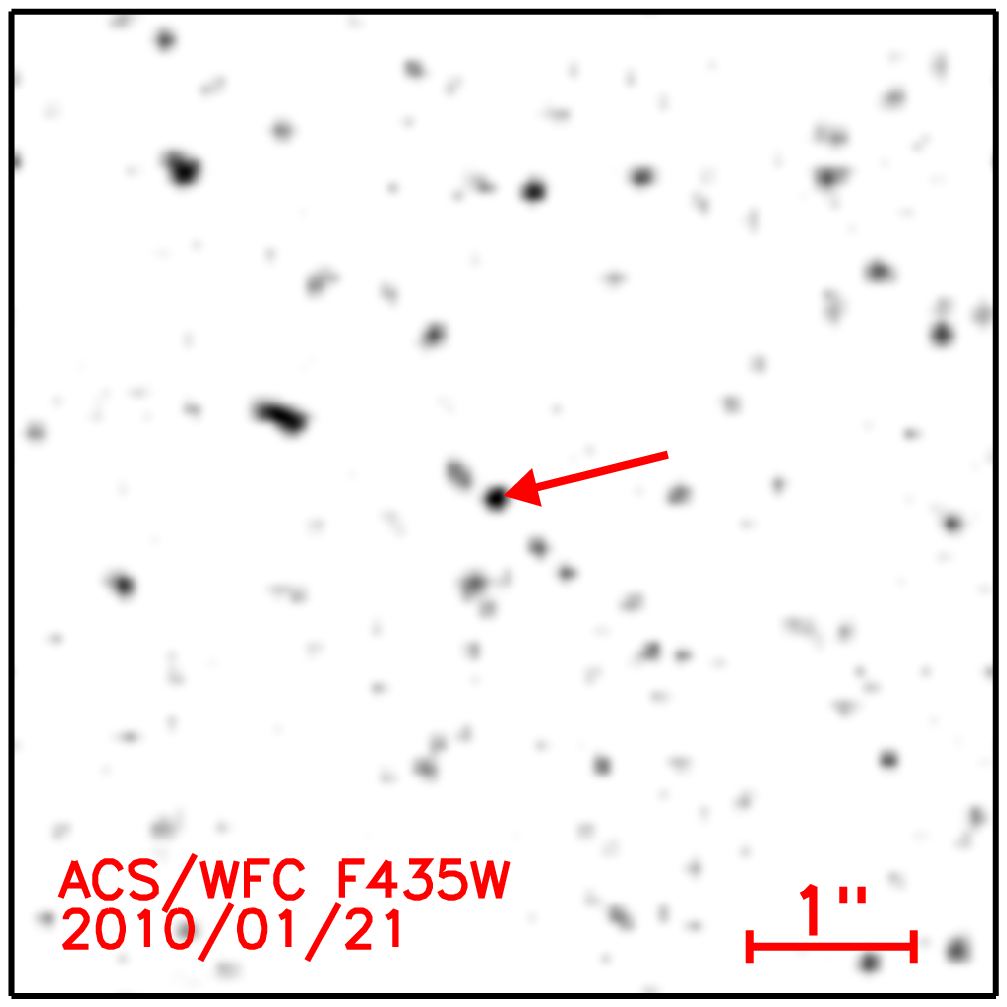}
\hspace{1cm}
 \epsfxsize=5cm \epsfbox{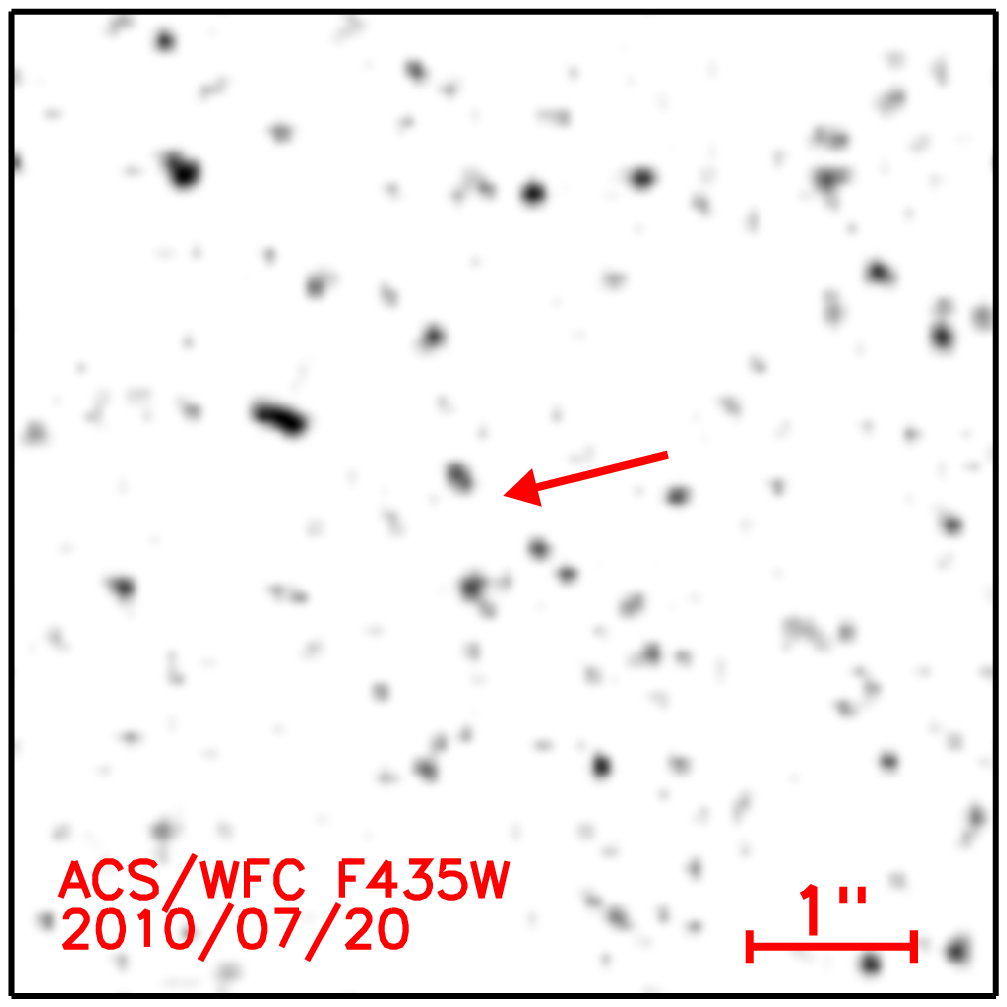}
\end{tabular}
\end{center}
\caption{HST (ACS) images taken on the 21st of Jan 2010 (left, OBSID:
JB9D15010, PropID=11833) and 20th of July 2010 (right, OBSID: JB9D20010,
PropID=11833) for exposures of 4360~s each. The highlighted position is that given in the finding charts of Garcia et al. (2010, ATEL \#2474) and,
whilst the former observation corresponds to an apparent optical
magnitude of 23.8 we determine the latter, quiescent observation to be
far fainter: $m_{\rm F435W} \ga 26$. Using the Galactic line-of-sight extinction (Predehl \& Schmitt 1995), we derive an upper limit on
the absolute magnitude of the optical counterpart in quiescence of
$M_{\rm F435W} \approx +1$, ruling out a high mass O or B star
companion to the compact object.  }
\label{fig:l}
\end{figure*}

The sequence of spectra during the decline can be very roughly
described by a standard disc model i.e. inner radius fixed at the
innermost stable circular orbit, around a stellar mass ($\sim
$10~M$_{\odot}$) BH.  If this is the correct interpretation then, for
the first time, we are observing a disc in the Eddington-regime
without the hindrance of a large neutral Galactic column, typical of
XRBs in our Galaxy. However, this gives residuals of order 10\% at low
and high energies. This could indicate that even our best disc models
are incomplete, e.g. they do not include advection or release of
advected radiation from the plunging region or self--shielding (S{\c
a}dowski 2009; S{\c a}dowski et al. 2009). However, these should all
give an increasing distortion of the disc spectrum at Eddington and
beyond, yet the residuals are seen at a similar or even greater extent
in the lowest Eddington fraction spectrum as well as in the highest
ones.  If this is the correct interpretation of the spectra then our
best current disc models can only describe the data within 10\% over
this wide bandpass. New disc models that fully incorporate the effects
of advection, relativistic smearing and energy transport will
identify, in a more physically robust manner, how current disc models
fail to describe the emission. This dataset will therefore prove
an invaluable measure of these future models' success (Straub et al. in prep).

Alternatively, the spectra are statistically better fit by an
advective (slim) disc model, together with a low temperature,
optically thick Comptonisation component which could be the
photosphere of an inner-disc wind (e.g. Middleton et al. 2011b). The
observed decrease in extent of this wind/photosphere as the luminosity
decreases is then due to the reduction in mass loss rate into the wind,
so that its photosphere covers less of the inner disc. The observed
energy-dependent variability (weakly) supports this two component
model, but we caution that this is close to the statistical limit of
what the variability data can probe.

We note that the situation remains partially degenerate in terms of
spectral components and viewing angles. We would obtain similar fit
statistics if the cool component was in fact an outer photosphere and
the hot region a modified `bare' disc where any loose material has
already been uplifted and removed in a wind. This is our proposed
model for NGC~5408 X-1 (Middleton et al. 2011a) and could likewise
predict the same presence of uncorrelated variability via extrinsic
means but at a different viewing angle. Whilst this is a much more
likely solution for the higher luminosity ULXs, we are currently unable
to break this degeneracy in lower luminosity ULXs such as this (and
M33 X-8; although that particular source has no constrained
variability: Middleton et al. 2011b). In either case, we require a two
component solution comprising a wind/photosphere and modified disc
emission. 

Hence we have a transient system that displays an extraordinary
combination of ULX-like luminosities, an {\it e}-fold time and derived
binary system parameters, and optical counterpart consistent with a
classical LMXB transient, and super-Eddington outburst behaviour
similar to other ULXs.  Additionally, the presence of an IMBH is
rejected on the basis of the X-ray characteristics, firmly
demonstrating that an IMBH is not necessary to reach ULX luminosities.
First and foremost, this then provides a solid evidential link between
accretion onto an apparent stellar mass black hole binary and ULX-like
behaviour.  This evidence exceeds that from previous analyses where
the success of super-Eddington emission models lead to the inference
of an underlying stellar-mass black hole (e.g. Gladstone et al. 2009);
in this case the transience allows far stronger physical links to be
made.

However, the transient nature of this object also marks it out as an
unusual ULX.  Although it has long been known that ULXs are associated
with both old and young stellar populations (Humphrey et al. 2003; Colbert
et al. 2004), the majority of work has focused on the objects directly
linked to young stellar populations as they are both more numerous and
more luminous than the older population (cf. Walton et al. 2011).  These
objects are also typically persistently luminous, unlike the transient
discussed here; in fact few transient ULXs have been observed, with
notable cases associated with the old population of Cen A (Ghosh et al.
2006; Sivakoff et al. 2008) and a recurrent transient in the outer regions of the archetypal
nuclear starburst galaxy NGC~253 (Bauer \& Pietsch 2005).  However,
despite the likely differences in donor star and outburst timescales, the
similarity in X-ray spectral characteristics between M31 ULX-1 and other
well-studied low luminosity ULXs (e.g. M33 X-8, Middleton et al. 2011b)
argues for a commonality in accretion physics and an underlying black hole
mass in the stellar regime in many ULXs.

\section{Acknowledgements}

The authors thank the anonymous referee for their useful comments. MM, TR and CD thank STFC for support in the form of a standard grant,
and AS similarly thanks STFC for support via a PhD studentship. This
work is based on observations obtained with {\it XMM-Newton}, an ESA
science mission with instruments and contributions directly funded by
ESA Member States and NASA. We also acknowledge use of archival {\it
Chandra} and {\it Swift} data. Additionally this work has made use of
observations made with the NASA/ESA Hubble Space Telescope, obtained
from the data archive at the Space Telescope Institute. STScI is
operated by the association of Universities for Research in Astronomy,
Inc. under the NASA contract NAS 5-26555

\label{lastpage}

\end{document}